# Resolving the Dirac Cone on the Surface of $Bi_2Te_3$ Topological Insulator Nanowires by Field-Effect Measurements

Johannes Gooth[1,*], Bacel Hamdou[1], August Dorn[1], Robert Zierold[1] and Kornelius Nielsch[1,#]

[1] Institute of Applied Physics, Universität Hamburg, Jungiusstrasse 11, 20355 Hamburg, Germany

Electronic address: [*] jgooth@physnet.uni-hamburg.de, [#] knielsch@physnet.uni-hamburg.de

ABSTRACT

**We validate the linear dispersion relation and resolve the Dirac cone on the surface of a single $Bi_2Te_3$ nanowire via a combination of field-effect and magnetoresistance measurements by which we unambiguously prove the topological insulator nature of the nanowire surface states. Moreover we show that the experimentally determined carrier concentration, mobility and cyclotron mass of the surface states are in excellent agreement with relativistic models. Our method provides a facile way to identify topological insulators that too small for angle-resolved photo emission spectroscopy.**

Unambiguous experimental identification of topological insulators (TIs) requires resolving the Dirac cone in the band structure of the non-trivial surface states.[1-4] Charge carriers within the surface states mimic massless Dirac fermions with extremely high mobilities, opening great opportunities for novel information processing devices. Nanowires with a high surface-to-volume ratio are of special interest because the surface state conductivity is expected to overwhelm the parasitic transport through the nanowire bulk.[5-10] While the Dirac cone has been directly detected by angle-resolved photoemission spectroscopy (ARPES) of $Bi_2Te_3$,[11-13] $Bi_2Se_3$,[12,14] $Sb_2Te_3$,[15] $TlBiSe_2$,[16] $Bi_{1-x}Sb_x$,[17,18] and $(Bi_{1-x}Sb_x)_2Te_3$[19] bulk samples and thin films, experimentally verifying the linear dispersion relation on the surface of TI nanowires has been proven elusive–a key challenge being the small lateral size of nanowires. Indications for two-dimensional surface states in nanowires are commonly obtained by electrical transport measurements, where phenomena such as Aharonov-Bohm (AB) oscillations,[5,6,8-10] the ambipolar field-effect,[20,21] universal conductance fluctuations (UCFs)[22] or weak antilocalization (WAL)[5,8-10,23,24] comprise the most compelling evidence for (two-dimensional) TI surface states to date. Furthermore, transport parameters which are consistent with massless Dirac fermions, such as very low cyclotron masses, high mobilities and Fermi velocities of around $4 \cdot 10^5$ ms$^{-1}$ are extracted from Shubnikov-de Haas (SdH) oscillations[5,7,8,21,25] measured in nanowires using a two-dimensional relativistic model. However, AB oscillations, the ambipolar field effect, UCFs, WAL and SdH oscillations have also been observed in non-topological insulator nanowires, resulting e.g. from two-dimensional charge accumulation surface layers,[26,27] narrow band gaps,[28,29] bulk confinement,[30,31] strong spin-orbit coupling,[32,33] or high bulk crystallinity.[7] Moreover, WAL, AB oscillations and SdH interferences are not immediately related to Dirac's (relativistic) equation but might also be accurately described by the non-relativistic Schrödinger equation.[34-36] WAL, the AB and the SdH effect are therefore not unambiguous evidence for the presence of topological surface states and resolving the Dirac cone is still necessary for a conclusive identification of TIs.

Here we validate the linear dispersion relation of the surface states of a $Bi_2Te_3$ nanowire by gate voltage-dependent magnetotransport measurements, proving the relativistic Dirac character of the surface states and thus the topological insulator nature of the nanowire. Our method provides a versatile tool to identify TIs that are inaccessible by ARPES measurements due to the small dimensions of the specimen. Moreover magnetotransport studies reveal additional information about charge transport in TIs, such as effective mass, carrier concentration and mobility. Specifically, we measured the resistance $R$ of a rectangular $Bi_2Te_3$ nanowire as a function of perpendicular magnetic field $B$ for various gate voltages $V_G$ at low temperatures $T$. Two distinct SdH oscillations–which we attribute to the top and the bottom surface of the nanowire–were observed. A quadratic correlation of the SdH frequency and the cyclotron mass was found by tuning the gate voltage. Further transport analysis reveals that this experimentally obtained dependence directly implies a linear relation between energy and $k$-wave vector.

Similar magnetotransport studies on gated TI nanostructures have been carried out previously for example on $Bi_2Te_3$ as well as on $Bi_2Se_3$ thin films,[37] nanoribbons,[5] and crystals,[38,39] where several groups have been able to distinguish two conducting surfaces.[20,37,38] While most of these works focus on reducing the relative contribution to electrical transport from the bulk by enhancing the surface contribution via field-effect, resolving the linear dispersion relation of the surface states has not been attempted so far.

Our experiments were performed using a single nanowire device equipped with two electrodes for resistance measurements and a global back gate (figure 1 a). Electrical contacts to the nanowire were prepared via common laser beam lithography and subsequent sputter deposition of Ti/Pt. A highly p-doped silicon chip served as gate with a 200 nm insulating buffer layer of thermal $SiO_2$ on top. Standard low frequency lock-in techniques were performed in at low temperatures with a magnetic field $B$ applied perpendicular to the current direction along the nanowire axis. Comparing two-terminal and four-terminal measurements reveal that the contact resistances to the nanowire are negligible. Linear IV-curves (SI Fig. 1) demonstrate that the electrical contacts ore ohmic. The Fermi energy of the nanowire was tuned by applying gate voltages between -10 and 30 V in 5 V steps, using a DC voltage source. $Bi_2Te_3$ nanowires were grown in a single zone tube furnace on $Si/SiO_2$ wafers seeded by 30 nm colloidal diameter Au particles, prior to 100 h annealing in a tellurium rich atmosphere at 250 °C. Further details on the nanowire growth as well as on electrical and on structural characterization can be found elsewhere.[8,40] We present data for a single-crystalline, rectangular $Bi_2Te_3$ nanowire with a cross-sectional area of (44 nm (height) x 200 nm (width)) and 1 μm channel length (figure 1 b). The nanowire was taken from a batch which had previously yielded nanowires with pronounced WAL, SdH interferences and AB oscillations allowing for direct validation of these broadly accepted indications for the presence of TI states.

The resistance of the $Bi_2Te_3$ nanowire investigated here shows a metallic temperature-dependence (SI Fig. 1) below room temperature, which is typically observed for $Sb_2Te_3$ nanowires[9], and other TI nanostructures, such as $Bi_2Se_3$ nanoribbons and thin films or $Bi_2Te_3$ crystals.[10, 41-43] Regardless of the gate voltage we observe three pronounced features in the magnetoresistance at 2 K as is exemplarily shown for $V_G = +10$ V in Fig. 1c: first, a smoothly varying background, ascribed to the nanowire bulk;[5,7,9] second, a sharp dip around zero magnetic field, attributed to WAL consistent with strong spin-orbit coupling;[5,8-10,23,24] and third, at higher magnetic fields, superimposed SdH oscillations that reveal a $1/B$ periodicity. For data analysis we will concentrate on the SdH contribution $\Delta R$, which we obtain by subtracting the smooth bulk background and the WAL contribution from $R$ (Fig. 1 d).

We observe two dominant transport channels in the SdH spectra by tuning the gate voltage (Fig. 2 a - channel I: light blue; and channel II: red, the Landau indices are enumerated for both channels), most likely related to topological charge transport on the top and on the bottom surface of the $Bi_2Te_3$ nanowire (Fig. 2 b).[20,37,38] For $1/B < 0.7$ T$^{-1}$ additional quantum features, which might originate from states or Universal Conductance Oscillations in the nanowire bulk,[9] does not allow for a clear identification of the lowest Landau levels at all gate voltages. For both SdH-channels a non-zero π-Berry phase is revealed in the intercepts of the Landau index fan diagram[44] (Fig. 3a and b). The non-zero π-Berry phase is usually considered to be a strong indication for TI surface states.[8, 43, 45] However, as is particularly evident for fits based on only a small number of oscillation periods, the intercepts scatter around the value predicted for the existence of Dirac particles: -0.5. For further analysis the SdH frequencies $B_F$ as well as the cyclotron mass $m_c$ of each transport channel are extracted at various gate voltages. $B_F$ is determined from the slope of the index plots and $m_c$ from the temperature damping (Fig. 1d) of the SdH conductance amplitudes $\Delta G$ that correspond to the first Landau level. Fitting the standard expression $\Delta G(T)/\Delta G(0) = \chi(T)/\sinh[\chi(T)]$ with $\chi(T) = \frac{2\pi^2 k_B T m_c}{\hbar e B}$ yields good agreement with the data (Fig. 3c and d). We obtain SdH frequencies in the range of 0.7 – 2.8 T and relatively low cyclotron masses of 0.013 – 0.024 $m_0$ in both transport channels. $B_F$ as well as $m_c$ increase with increasing gate voltage. Note, that so far we have made no assumptions about the energy dispersion in momentum space, i. e. the band structure.

When relating the SdH frequency $B_F$ and the cyclotron mass $m_c$ at each gate voltage we find a quadratic correlation of the SdH frequency and the cyclotron mass (Fig. 3e and f). $B_F \sim m_c^2$ is obtained from linear fits in the double logarithmic plots (SI Fig. 2). This experimentally determined dependence directly implies a linear relation between energy and momentum, which validates the relativistic Dirac character of the two observed transport channels. Our line of argumentation follows a method applied to graphene,[46] with the advantage that solely magnetoresistance but no supplementary Hall measurements are required: Combining the semi-classical expressions $B_F = (\hbar/2\pi e)S(E)$ and $m_c = (\hbar^2/2\pi)\partial S(E)/\partial E$, where $S(E) = \pi k^2$ is the area of orbits at the Fermi energy $E(k)$ in momentum space, with the experimentally found relationship $B_F \sim m_c^2$, it is straightforward to show that $k \sim \frac{\partial}{\partial E}k^2$, which directly yields $E \sim k$. Note that the model is valid for the Schrödinger as well as for the Dirac case, fully justified by a recent theory.[47, 48] Our gate voltage-dependent data therefore unambiguously demonstrate the existence of the relativistic Dirac particles with linear dispersion relation $E = \hbar v_F k$ in $Bi_2Te_3$ nanowires as expected for topological surface states.[1-4] The non-zero $\pi$-Berry phase as well as the low cyclotron masses are consistent with this result. The best fit to our data, converted in momentum $k$ and energy $E$ by $k = (2\pi B_F/\hbar)^{1/2}$ and $E = (\hbar k)^2/m_c$ (Fig. 4 a), results in a Fermi velocity of $v_F = 4.35 \cdot 10^5$ ms$^{-1}$ in channel I and $v_F = 4.25 \cdot 10^5$ ms$^{-1}$ in channel II, which is in good agreement with ARPES measurements[11-13] and band structure calculations.[3] Furthermore, the carrier concentration $n_s$ in the two channels can be estimated from $B_F$ using the expression[5,7,8,45] $n_s = k^2/2\pi$. On this basis we find that the gate-tuned cyclotron mass, shown in Fig. 4 c and d, traces the square-root dependence of the fictitious relativistic mass[44] $m_c^2 = E_F/v_F^2 = (\pi \hbar^2 n_s/v_F^2)^{1/2}$ very well.

Since TIs exhibit only one non-degenerated Dirac cone at each surface, the two assigned transport channels must belong to two different surfaces of the nanowire that are perpendicular to the applied magnetic field. More specifically, we attribute channel I to the top surface and channel II to the bottom surface of the nanowire. Channel II is probably located at the interface to the $SiO_2$ substrate (schematically shown in Fig. 2b), based on two experimental observations: first, the gate coupling $dE/dV_G$ to channel I ($dE/dV_G$ = 230 µeV/V) is weaker than to channel II ($dE/dV_G$ = 252 µeV/V); and second, extracting the electron mobility $\mu$ from the magnetic field (Dingle) damping of the SdH oscillation amplitude[5,7,8] $\chi(T)/\sinh[\chi(T)]e^{-\pi/\mu B}$ (see SI Fig. 3 for details), we obtain a higher mobility in channel I ($\mu \approx$ 45 000 cm$^2$V$^{-1}$s$^{-1}$) than in channel II ($\mu \approx$ 38 000 cm$^2$V$^{-1}$s$^{-1}$) for a similar carrier concentration

($n = 10^{11}$ cm$^{-2}$) at 2 K. Weaker gate coupling to the top surface as compared to the bottom surface channel of the nanowire is expected because of the larger distance to the gate electrode and as a result of screening by the bottom channel. The mobility at the nanowire-SiO$_2$ interface might be suppressed compared to the nanowire-vacuum interface due to interaction of the Dirac fermions with impurities and trap states at the SiO$_2$ surface.

We have demonstrated the existence of a Dirac-like linear dispersion relation $E = \hbar v_F k$ for the surface states on a Bi$_2$Te$_3$ nanowire. Our experiments can be understood as an energy scan across the spectrum of the top and of the bottom surface of the nanowire. The Fermi energy is controlled by a gate voltage and characteristic transport parameters are simultaneously probed by magnetoresistance measurements. The energy spectrum reveals a Dirac cone, which implies a vanishing cyclotron mass and carrier concentration near the Dirac point (E = 0), closely related to the observed π-Berry phase. Moreover we showed that the experimentally determined cyclotron mass of the surface states is well described by the relativistic equation $m_c^2 = (\pi \hbar^2 n_s / v_F^2)^{1/2}$. The good agreement between our evaluated transport parameters and other reports, in which the linear dispersion relation had not been determined, supports the common indicators used to identify topological surface states in nanowires–AB, SdH oscillations and the WAL effect. However, only resolving the Dirac cone at the nanowires' surface allows for the unambiguous identification of a TI. Electric field-dependent magnetotransport studies present a facile, alternative way to identify the Dirac cone and thus prove the TI nature of nanostructures that are inaccessible by ARPES measurements.

ACKNOWLEDGEMENTS

We thank Ulrich Merkt, Martin Waleczek and Phillip Sergelius for useful discussions. This work was supported by the Deutsche Forschungsgemeinschaft (DFG) via Graduiertenkolleg 1286 "Functional Metal-Semiconductor Hybrid Systems" as well as SPP 1386 and SPP 1666.

FIGURES

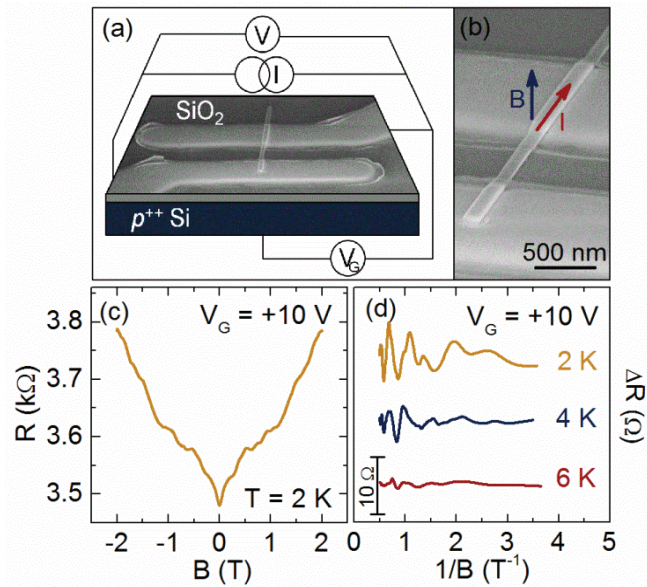

**FIG. 1.** Electric field-dependent magnetoresistance measurements on a single $Bi_2Te_3$ nanowire at low temperatures. (a) Scheme of the measurement device based on a scanning electron microscope (SEM) image of the nanowire and two contact leads. The highly *p*-doped $Si/SiO_2$ chip (blue/dark grey) serves as a global back gate. The resistance of the nanowire is measured with standard low frequency lock-in technique. (b) (SEM) image of the investigated nanowire with a rectangular cross-sectional area of (44 (height) x 200 (width)) $nm^2$ and 1 µm channel length. The magnetic field $B$ is applied perpendicular to the $SiO_2$ surface as well as to the current direction along the nanowire axis. (c) Typical magnetoresistance curve exemplarily shown for a back-gate voltage of $V_G$ = +10 V at a substrate temperature of 2 K. Three feature can be identified: a weak antilocalization (WAL) peak around $B$ = 0 T; a smooth varying background as well as superimposed Shubnikov de Haas (SdH) oscillations that are periodic in $1/B$. (d) The SdH modulations become more pronounced at low temperatures, which can be clearly seen in $\Delta R(T)$ at different temperatures $T$ (6 K: red, 4 K: yellow, 2 K: dark blue) after subtraction of the smooth background and the WAL contribution from $R$.

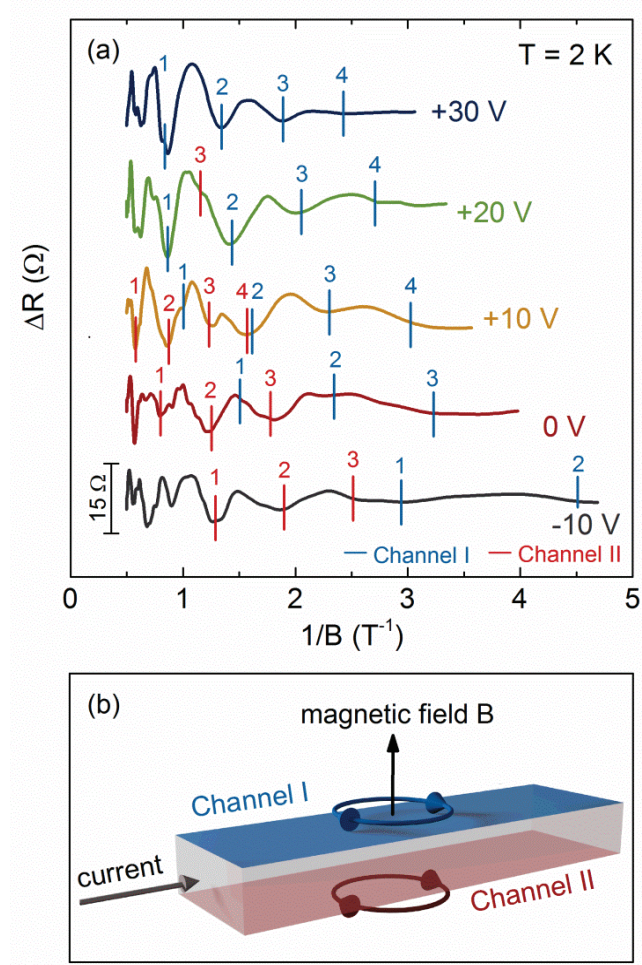

**FIG. 2.** Shubnikov-de Haas oscillations in two transport channels located at the top and at the bottom surface of the $Bi_2Te_3$ nanowire (channel I: light blue, top; channel II: light red, bottom). (a) Shubnikov-de Haas contribution to the total electrical resistance $\Delta R$ as a function of inverse magnetic field for various gate voltages (+30 V: dark blue, +20 V: green, +10 V: yellow, 0 V: dark red, -10 V: dark grey) at 2 K. Distinct SdH frequencies are observed attributed to two transport channels. The individual Landau levels are enumerated for each channel with the corresponding color. (b) Scheme of the individual transport channels (light blue: upper surface; light red: bottom channel).

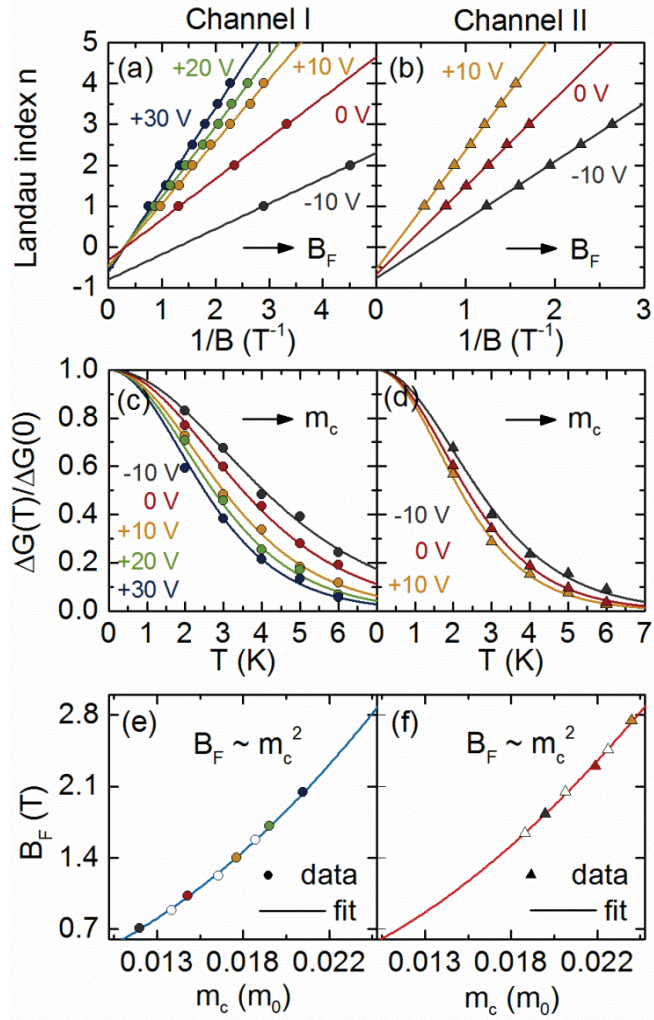

**FIG. 3.** Quadratic correlation of the SdH frequency on the cyclotron mass observed in a Bi$_2$Te$_3$ nanowire for the two individual channels. (a),(b) Experimentally determined Landau level $n$ (symbols) versus inverse magnetic field $1/B$ for various gate voltages $V_G$ (+30 V: dark blue, +20 V: green, +10 V: yellow, 0 V: dark red, -10 V: dark grey) revealing the SdH frequency $B_F$ from the slope of linear fits (solid lines) and the $\pi$-Berry phase from the extrapolated intercepts. In addition to integer values of $n$, which are defined by resistance minima, the half integer maxima are plotted for precise fitting. (c),(d) Experimentally determined temperature dependence of the SdH oscillation conductance amplitude at $n = 1$ (symbols) fitted by $\Delta G(T)/\Delta G(0) = \chi(T)/\sinh[\chi(T)]$ with $\chi(T) = \frac{2\pi^2 k_B T m_c}{\hbar e B}$ (solid lines) to extract the cyclotron mass $m_c$ at various $V_G$ (same color-code as in (a) and (b)). (e),(f) Squared dependence of the

SdH frequency on the cyclotron mass (symbols)—extracted from (a),(b),(c) and (d)—from which it is straightforward to show that $E \sim k$. The corresponding gate voltages are color-coded equivalently to (a),(b),(c) and (d). The white symbols mark gate-voltages at which $n(1/B)$ and $\Delta G(T)/\Delta G(0)$ are measured, but not shown in here. The quadratic correlation between $B_F$ and $m_c$ is determined from linear fits of logarithmic plots of $B_F$ and $m_c$ (SI figure 2) at various gate voltages, shown as solid lines with $B_F \sim m_c^{2.07}$ in channel I and $B_F \sim m_c^{1.96}$ in channel II.

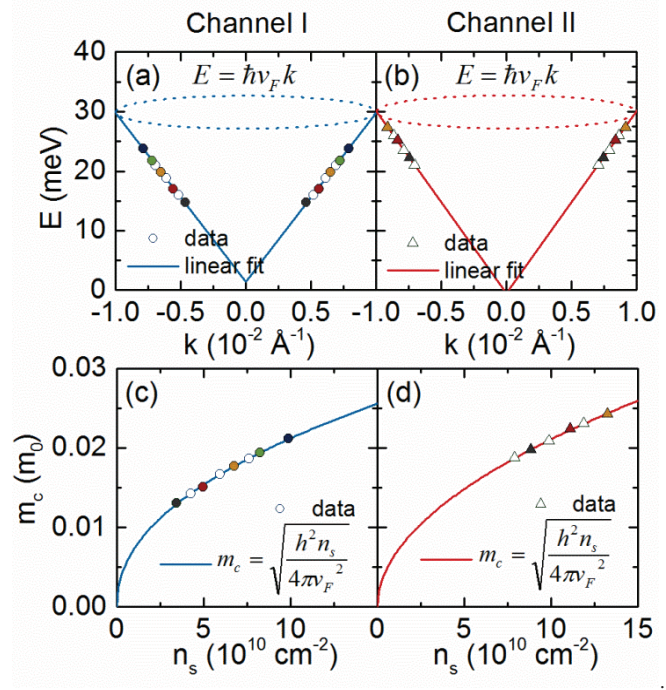

**FIG. 4.** Dirac cone in the band structure of the top (left) and the bottom (right) surface of the Bi$_2$Te$_3$ nanowire. (a),(b) The linear dispersion relation at each surface proves the existence of the relativistic Dirac particles with linear dispersion relation $E = \hbar v_F k$ in Bi$_2$Te$_3$ nanowires as expected for topological surface states. The corresponding gate-voltages are equally color-coded as in figure 2 (+30 V: dark blue, +20 V: green, +10 V: yellow, 0 V: dark red, -10 V: dark grey). Best fits to our data results in a Fermi velocity of $v_F = 4.35 \cdot 10^5$ ms$^{-1}$ and $v_F = 4.25 \cdot 10^5$ ms$^{-1}$ at the top surface (channel I) and at the bottom surface (channel II), respectively. (c),(d) The gate-tuned cyclotron mass $m_c$ as a function of carrier concentration $n_s$ traces the square-root dependence of the fictitious relativistic mass $m_c^2 = E_F/v_F^2 = (\pi\hbar^2 n_s/v_F^2)^{1/2}$ very well.

**Supporting information for the manuscript**

*Resolving the Dirac Cone on the Surface of Bi$_2$Te$_3$ Topological Insulator Nanowires by Field-Effect Measurements*

By Johannes Gooth, Bacel Hamdou, August Dorn, Robert Zierold and Kornelius Nielsch

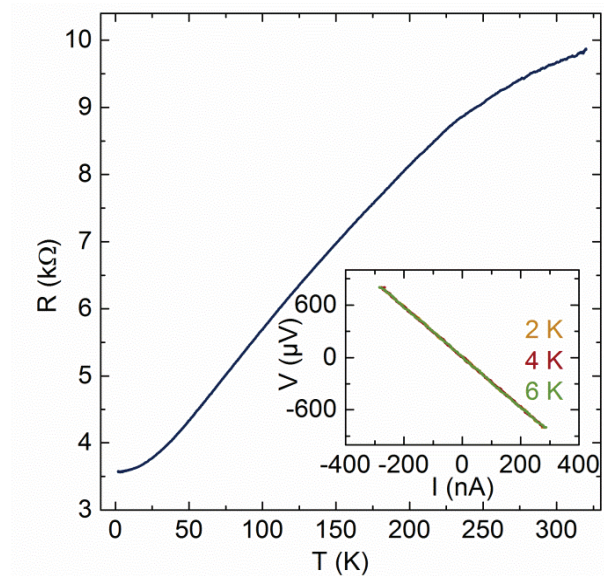

**SI FIG 1.** Temperature-dependent resistance of the investigated Bi$_2$Te$_3$ nanowire with rectangular cross-sectional area of (44 nm (height) x 200 nm (width)) and 1 μm channel length shows typical metallic behavior. The measurement is done with no magnetic field applied between 300 and 2 K. The inset shows IV-curves at temperatures of 2 K (yellow), 4 K (red) and 6 K (green), covering the temperature range in which the magnetoresistance analysis of the manuscript has been performed. The linearity of the IV-diagrams indicates ohmic contacts.

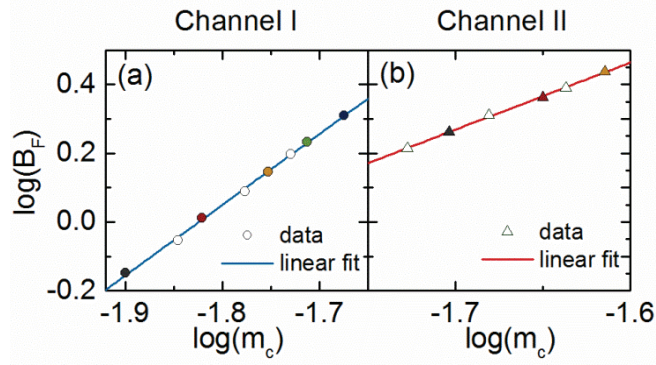

**SI FIG 2.** Double logarithmic plot of the Shubnikov-de Haas frequency $B_F$ versus the cyclotron mass $m_c$ for (a) channel I and (b) transport channel II (left and right column of the figure) of the investigated $Bi_2Te_3$ nanowire. The symbols correspond to data taken from figure 2 (a), (b), (c) and (d) of various gate voltages. The symbols are color-coded equivalently to figure 2 (+30 V: dark blue, +20 V: green, +10 V: yellow, 0 V: dark red, -10 V: dark grey). The white symbols mark gate-voltages at which the Landau indices are measured as a function of inverse magnetic field $n(1/B)$ and the conductance as a function of temperature $\Delta G(T)/\Delta G(0)$ (not shown in the manuscript). Linear fits (solid lines) to the data reveal $\log(B_F)$ = -1.453 (± 0.002) + 0.482 (± 0.008) $\log(m_c)$ for channel I and $\log(B_F)$ = -1.837 (± 0.004) + 0.511 (± 0.001) $\log(m_c)$ for channel II. From this we find $B_F \sim m_c^{2.07}$ in channel I and $B_F \sim m_c^{1.96}$ in channel II—a squared dependence between $B_F$ and $m_c$—from which it is straightforward to show that the Fermi energy $E$ is directly proportional to the k-wavevector.

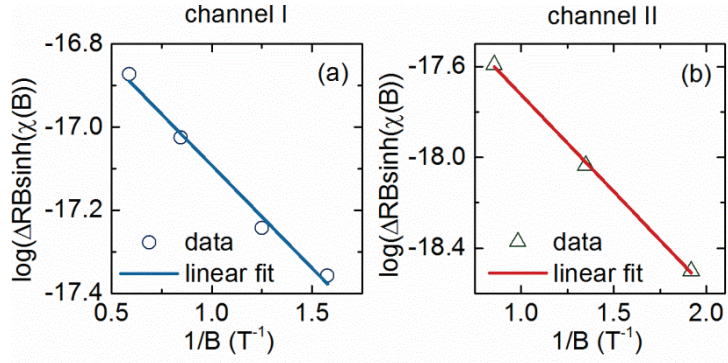

**SI FIG 3.** Carrier mobility $\mu$ at the (a) top (channel I) and at the (b) bottom (channel II) surface of the investigated $Bi_2Te_3$ nanowire for a carrier concentration of $n = 10^{11}$ cm$^{-2}$ at $T = 2$ K base temperature. $\mu$ is calculated from the Dingle damping factor of disorder[1,2] $R_D = e^{-\pi/\tau\omega}$ where $\tau$ is the scattering time and $\omega$ the cyclotron frequency. The resistance amplitude of the Shubnikov-de Haas oscillations is proportional to $\chi(B)/\sinh[\chi(B)]e^{-\pi/\mu B}$ with $\mu = \tau\omega/B$ and $\chi(B) = \frac{2\pi^2 k_B T m_c}{\hbar e B}$, where $B$ is the magnetic field and $m_c = 0.02\, m_0$ the cyclotron mass at $n = 10^{11}$ cm$^{-2}$, corresponding to a gate voltage $V_G = +30$ V applied to channel I and $V_G = -10$ V applied to channel II. When plotting $RB\sinh[\chi(T)]$ against $1/B$, the carrier mobility can be extracted from the slope of the logarithmic plots.[3-5] We obtain a higher mobility in channel I ($\mu \approx 45\,000$ cm$^2$V$^{-1}$s$^{-1}$) than in channel II ($\mu \approx 38\,000$ cm$^2$V$^{-1}$s$^{-1}$) for similar carrier concentrations, which is likely due to interaction of the Dirac fermions at the bottom surface with impurities in the $SiO_2$ substrate.